\documentclass[pre,preprint,showpacs,eqsecnum]{revtex4}
\usepackage{graphicx}
\begin{document}
\title{Anomalous biased diffusion in a randomly layered medium}
\author{S.~I.~Denisov$^{1,2}$}
\email{stdenis@pks.mpg.de}
\author{H.~Kantz$^{1}$}
\affiliation{$^{1}$Max-Planck-Institut f\"{u}r Physik komplexer Systeme,
N\"{o}thnitzer Stra{\ss}e 38, D-01187 Dresden, Germany\\
$^{2}$Sumy State University, 2 Rimsky-Korsakov Street, UA-40007 Sumy, Ukraine }


\begin{abstract}
We present analytical results for the biased diffusion of particles moving
under a constant force in a randomly layered medium. The influence of this
medium on the particle dynamics is modeled by a piecewise constant random
force. The long-time behavior of the particle position is studied in the frame
of a continuous-time random walk on a semi-infinite one-dimensional lattice. We
formulate the conditions for anomalous diffusion, derive the diffusion laws and
analyze their dependence on the particle mass and the distribution of the
random force.
\end{abstract}

\maketitle

\section{INTRODUCTION}

A vast variety of physical, chemical, biological and other natural processes
can be adequately described by random processes exhibiting anomalous diffusion
behavior at long times. This behavior, which is characterized by a nonlinear
dependence of the variance of these processes on time, can be observed in
various systems. Anomalous diffusion actually occurs, for example, in turbulent
fluids \cite{Rich}, amorphous solids \cite{SM}, rotating flows \cite{SWS},
single molecules \cite{YLK} and porous substrates \cite{BJB}, and has been
predicted to occur in many other systems \cite{BG,AH,MK,Z,KRS}.

The existence of anomalous diffusion in systems with quenched, i.e.,
time-independent disorder has also been extensively studied \cite{BG,AH}. One
of the most effective and simple ways to describe anomalous diffusion in these
systems is based on the motion equations for diffusing objects (which we will
call particles). In these equations, the influence of quenched disorder is
usually modeled by a time-independent random potential producing the
corresponding random force, and the influence of thermal fluctuations is
accounted for by white noise. Specifically, this Langevin-type approach has
been successfully applied to study a variety of phenomena, including biased
diffusion, which occur when particles move under a constant external force in a
one-dimensional potential \cite{Const}.

If thermal fluctuations are absent then particles can be transported to an
arbitrary large distance only if the distribution of the random force has
bounded support. In this case, the directional transport of particles can be
caused by either a periodic external force \cite{Rat} or a constant one. In the
latter case, particles move only in one direction and so the completely
anisotropic case of biased diffusion, when the probability of motion along and
against the external force equals 1 and 0, respectively, may exist. It has been
shown for particular cases of the random force distribution that in the
overdamped limit this diffusion is normal if the total force acting on a
particle is strictly positive or strictly negative \cite{KLS, DKDH}. In
Ref.~\cite{KLS} it was also argued that the anomalous regimes of biased
diffusion would exist if the lower (upper) bound of the total force at a fixed
external force is equal to zero. However, none of the laws of anomalous
diffusion was found in this case.

The aim of this paper is to study the anomalous regimes of biased diffusion of
particles moving under a constant force in a randomly layered medium which acts
as a piecewise constant random force. The paper is organized as follows. In
Sec.~\ref{sec:Mod}, we describe the model, reduce it to a continuous-time
random walk (CTRW) on a semi-infinite chain, and calculate the first two
moments of the particle position. The connection between the waiting time
probability density and the particle mass and the probability density of the
random force is also presented in this section. The conditions providing the
anomalous behavior of biased diffusion are formulated in Sec.~\ref{sec:Cond}.
In Sec.~\ref{sec:Laws}, using the Tauberian theorem and its modified version,
we derive the laws of anomalous diffusion and analyze the influence of the
particle mass and the random force distribution. Finally, in
Sec.~\ref{sec:Concl} we summarize our results.

\section{MODEL AND BASIC EQUATIONS}
\label{sec:Mod}

\begin{figure}
    \centering
    \includegraphics[totalheight=4.5cm]{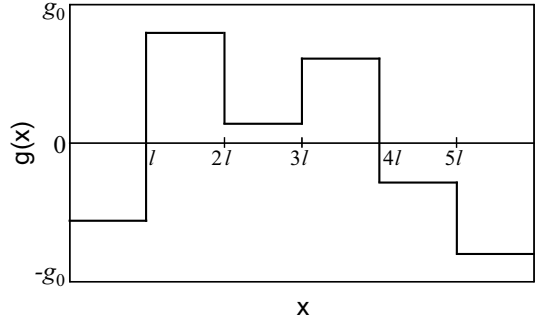}
    \caption{\label{fig.1} Sample path of a piecewise constant
    random force $g(x)$.}
\end{figure}

We consider the one-dimen\-sio\-nal propagation of a particle in a medium
composed by the layers of a fixed width $l$ whose transport properties are
assumed to be random. The motion of a particle in this medium occurs under the
action of a constant external force $f \,(>0)$, and the influence of the layers
on the particle dynamics is modeled by a random force $g(x)$. We assume that
$g(x)$ (i) is a bounded function, i.e., $g(x) \in [-g_{0}, g_{0}]$, (ii)
possesses a symmetry property, i.e., $g(x)$ and $-g(x)$ are statistically
equivalent, and (iii) has statistically independent values on different
intervals of the length $l$. In accordance with these conditions, we
approximate $g(x)$ by a piecewise constant random force (see Fig.~\ref{fig.1})
whose values are distributed with the same probability density $u(g)$. In this
stage, we consider $u(g)$ as an arbitrary symmetric probability density,
$u(g)=u(-g)$, satisfying only the normalization condition $\int_ {-g_{0}}^
{g_{0}} dg\,u(g) = 1$.

Since the total force acting on a particle equals $f + g(X_{t})$, where $X_{t}$
($X_{0}=\dot{X}_{0}=0$) is the particle position, its dynamics can be described
by the motion equation
\begin{equation}
    \mu\ddot{X}_{t} + \nu \dot{X}_{t} = f + g(X_{t})
\label{eq motion}
\end{equation}
with $\mu$ and $\nu$ being the particle mass and the damping coefficient,
respectively. According to this equation, if $f>g_{0}$ then $f+g(X_{t})>0$ and
particles can be transported to an arbitrary large distance in the positive
direction of the axis $x$. Thus, in this case the condition ${X_{t}|}_{t\to
\infty} \to \infty$ holds for all sample paths of $g(x)$. On the contrary, if
$f<g_{0}$ then for each sample path of $g(x)$ there always exists a certain
point $L=l n_{\text{st}}$ ($n_{\text{st}}=n$, $n = 0,1,\ldots$), which is
characterized by the conditions $f>-g(x)|_{x<L}$ and $f<-g(x)|_{x=L+0}$, where
particles are stopped. The probability $W_{n}$ that $n_{\text{st}}=n$ is
expressed through the probability $I=\int_{-g_{0} }^{f}dg\,u(g)$ that $g(x)<f$
as follows: $W_{n} = I^{n} - I^{n+1}$. Therefore, the average distance $\langle
L \rangle$ (the angular brackets denote an average over the sample paths of
$g(x)$) from the origin to the stopping point can be written in the form
$\langle L \rangle = l\sum_{n=1}^{\infty} nW_{n} = l\sum_{n=1}^{\infty}I^{n}$.
Finally, using the geometric series formula, we obtain the desired result
\begin{equation}
    \langle L \rangle = l\,\frac{I}{1-I}.
    \label{aver}
\end{equation}

Assuming that the probability that $|g(x)| =g_{0}$ equals zero, i.e., the
probability density $u(g)$ is not concentrated at the edges of the interval
$[-g_{0},g_{0}]$, one can make sure that $I \to 1$ and so $\langle L \rangle
\to \infty$ as $f\to g_{0}$. Hence, at $f = g_{0}$ the condition ${X_{t}
|}_{t\to \infty} \to \infty$ holds almost surely, i.e., with probability one.
In contrast, if the probability density $u(g)$ has unbounded support with
$g_{0}= \infty$ as, e.g., for a Gaussian distribution, then $I<1$ and so
$\langle L \rangle$ is finite for all finite values of the driving force $f$.
In other words, in this case particles cannot be transported to an arbitrary
large distance. It is therefore we consider here only a class of probability
densities $u(g)$ with bounded support. It should be noted in this context that,
since infinite values of $g(x)$ are physically not relevant, the assumption of
bounded support is not too restrictive.

Our aim is to study the long-time behavior of the particle position $X_{t}$ at
$f\geq g_{0}$. The main statistical characteristic of $X_{t}$ is its
probability density function $P(x,t)$ defined as $P(x,t) = \langle
\delta(x-X_{t}) \rangle$, where $\delta(x)$ is the Dirac $\delta$ function. If
the solution of Eq.~(\ref{eq motion}) were known for all sample paths of
$g(x)$, it would be, in principle, possible to determine $P(x,t)$ directly from
the definition. However, this approach is difficult to implement and, what is
more important, it is not necessary for finding the long-time behavior of the
moments of $X_{t}$. Moreover, since at long times $X_{t}$ can be accurately
evaluated as a total length of the intervals $(nl, nl+l)$ ($n=0,1, \ldots$)
which a particle passes, many of the details of the particle dynamics described
by Eq.~(\ref{eq motion}) are needless for this purpose.

It is therefore reasonable to consider, instead of the model based on
Eq.~(\ref{eq motion}), the unidirectional CTRW of a particle on a semi-infinite
one-dimensional lattice with the period $l$. Introduced more than four decades
ago \cite{MW}, the CTRW model has become one of the most effective and powerful
tools in the theory of anomalous diffusion (see, e.g., Refs.~\cite{MK,Z,KRS}).
Within this model, we describe the particle position by a discrete variable
$Y_{t} =l N(t)$, where $N(t)$ is the random number of jumps up to time $t$. In
order to guarantee that the long-time behavior of $Y_{t}$ and $X_{t}$ are the
same, we assume that for all $n$ the waiting time $\tau^{(n)}$, i.e., the time
of occupation of the site $nl$, is equal to the time that a particle spends
moving from the site $nl$ to the site $nl +l$. If the inertial effects can be
neglected then Eq.~(\ref{eq motion}) yields $\tau^{(n)} = \nu l/(f+ g^{(n)})$,
where $g^{(n)} =g(x)$ and $x$ belongs to the $n$th interval, i.e., $x\in [nl,
nl+l)$. Since the random forces $g_{n}$ are statistically independent and
distributed with the same probability density $u(g)$, the waiting times
$\tau^{(n)}$ are also statistically independent variables whose probability
density is given by
\begin{equation}
    p(\tau) = \left\{ \begin{array}{ll}
    \displaystyle \frac{\nu l}{\tau^{2}}\,
    u\!\left(\frac{\nu l}{\tau} -f\right),
    \quad \tau\in [\tau_{\rm{min}},\tau_{\rm{max}}]
    \\ [12pt]
    0, \quad \rm{otherwise}
    \end{array}
    \right.,
    \label{p}
\end{equation}
where
\begin{equation}
    \tau_{\rm{min}} = \frac{\nu l}{f+g_{0}}, \quad
    \tau_{\rm{max}} = \frac{\nu l}{f-g_{0}}.
    \label{tau}
\end{equation}

It is important to emphasize that the inertial effects, at least in the
underdamped regime characterized by the condition $\nu \tau_{\rm {min}}/\mu \gg
1$ (weakly underdamped regime), can also be incorporated into the CTRW
framework. In order to illustrate this, let us first write the particle
velocity $v^{(n)}(\Delta t)$ [$\Delta t \in (0, \tau^{(n)})$] on the $n$th ($n
\geq 1$) interval. The straightforward integration of Eq.~(\ref{eq motion})
yields
\begin{equation}
    v^{(n)}(\Delta t) = \frac{f+g^{(n)}}{\nu} + \left( v^{(n)}_{-} -
    \frac{f+g^{(n)}}{\nu} \right)\! e^{-\kappa \Delta t},
    \label{vn}
\end{equation}
where $\kappa = \nu/\mu$ and $v^{(n)}_{-} = v^{(n)}(0)$ is the particle
velocity at the left end of the $n$th interval. Introducing also the particle
velocity at the right end of this interval, $v^{(n)}_{+} = v^{(n)} (\tau^{
(n)})$, from Eq.~(\ref{vn}) we obtain
\begin{equation}
    v^{(n)}_{+} = \frac{f+g^{(n)}}{\nu} + \left( v^{(n)}_{-} -
    \frac{f+g^{(n)}}{\nu} \right)\! e^{-\kappa \tau^{(n)}}.
    \label{vn+}
\end{equation}
Then, taking into account that $l=\int_{0}^ {\tau^{(n)}} d(\Delta t)\, v^{(n)}
(\Delta t)$, with the help of Eqs.~(\ref{vn}) and (\ref{vn+}) we find
\begin{equation}
    l = \frac{f+g^{(n)}}{\nu}\,\tau^{(n)} - \frac{v^{(n)}_{+} -
    v^{(n)}_{-}}{\kappa}.
    \label{l}
\end{equation}

Since in the case under consideration $\kappa \tau^{(n)} \gg 1$, the
exponential term in Eq.~(\ref{vn+}) can be neglected yielding $v^{(n)}_{+} =
(f+g^{(n)})/\nu$. According to this approximation, the particle velocity at the
end of the $n$th interval is determined by the random force on this interval.
Therefore, using the continuity condition for the particle velocity,
$v^{(n)}_{-} = v^{(n-1)}_{+}$, we obtain $v^{(n)}_{-} = (f+g^{(n-1)} )/\nu$.
Substituting these expressions for $v^{(n)}_{+}$ and $v^{(n)}_{-}$ into
Eq.~(\ref{l}), we arrive to the following result:
\begin{equation}
    \tau^{(n)} = \frac{\kappa\nu l + g^{(n)} - g^{(n-1)}}
    {\kappa (f+g^{(n)})}.
    \label{taun}
\end{equation}
It shows that in the weakly underdamped regime the waiting time $\tau^{(n)}$
depends not only on the random force $g^{(n)}$, as in the overdamped case, but
also on the random force $g^{(n-1)}$. Since these forces are statistically
independent, the probability density of the waiting time can be written in the
form
\begin{equation}
    p(\tau) = \int_{-g_{0}}^{g_{0}}\int_{-g_{0}}^{g_{0}} dgdg'
    u(g)u(g')\, \delta\! \left( \tau - \frac{\kappa\nu l +g -
    g'}{\kappa (f+g)} \right)
    \label{p2}
\end{equation}
if $\tau \in [\tau_{\rm{min}}, \tau_{\rm{max}}]$, otherwise it equals zero. It
is not difficult to verify that in the overdamped case (when $\kappa = \infty$)
Eq.~(\ref{p2}) reduces to Eq.~(\ref{p}).

Next, we express the first two moments of the random variable $Y_{t}$ through
the waiting time probability density $p(\tau)$. Since the moments of $N(t)$ are
known from the CTRW theory (see, e.g., Ref.~\cite{Hughes}), we reproduce here
only the main results related to our situation. Introducing the probability
$\mathcal{P}(n,t)$ that $N(t)=n$, we define the $k$th moment of the particle
position $Y_{t}$ in the usual way:
\begin{equation}
    \langle Y^{k}_{t} \rangle = l^{k} \sum_{n=1}^{\infty}n^{k}
    \mathcal{P}(n,t)
    \label{kth}
\end{equation}
$(k=1,2,\ldots)$. Then, using the Laplace transform of a function $h(t)$,
$h_{s} = \mathcal{L}\{ h(t) \} = \int_{0}^{\infty}dt\, e^{-st}h(t)$
($\rm{Re}\,s>0$), and taking into account that $\mathcal{P}(0,t)
=\int_{t}^{\infty} d\tau\, p(\tau)$ and
\begin{equation}
    \mathcal{P}(n,t)=\int_{0}^{t}d\tau\, p(\tau)\mathcal{P}(n-1,t-\tau)
    \label{P(n,t)}
\end{equation}
($n\geq 1$), we obtain
\begin{equation}
    \langle Y^{k}_{t} \rangle_{s} = l^{k}\frac{1-p_{s}}{s} \left( p_{s}
    \frac{d}{dp_{s}} \right)^{\!k} \frac{1}{1-p_{s}}.
    \label{Lapl kth}
\end{equation}
Finally, applying to Eq.~(\ref{Lapl kth}) the inverse Laplace transform defined
as $h(t) = \mathcal{L}^{-1}\{ h_{s} \} = (1/2\pi i) \int_{c- i\infty}^
{c+i\infty} ds\,e^{st}h_{s}$ ($c$ is chosen to be larger than the real parts of
all singularities of $h_{s}$), we find the first
\begin{equation}
    \langle Y_{t} \rangle = l\,\mathcal{L}^{-1}\!\left\{ \frac{p_{s}}
    {s(1-p_{s})} \right\}
    \label{1st}
\end{equation}
and the second
\begin{equation}
    \langle Y^{2}_{t} \rangle = l^{2}\mathcal{L}^{-1}\!\left\{
    \frac{p_{s}^{2} + p_{s}}{s(1-p_{s})^{2}} \right\}
    \label{2nd}
\end{equation}
moments of $Y_{t}$, which in turn determine the variance of the particle
position:
\begin{equation}
    \sigma^{2}(t) = \langle Y^{2}_{t} \rangle - \langle Y_{t} \rangle^{2}.
    \label{def var}
\end{equation}

\section{CONDITIONS OF ANOMALOUS DIFFUSION}
\label{sec:Cond}

As it follows from the waiting time probability density (\ref{p2}), the $m$th
moment of the waiting time, $\overline{\tau^{m}} = \int_{0}^{\infty}
d\tau\,\tau^{m} p(\tau)$ $(m=1,2,\ldots)$, can be written in the form
\begin{equation}
    \overline{\tau^{m}} = \int_{-g_{0}}^{g_{0}}\int_{-g_{0}}^{g_{0}}
    dgdg' u(g)u(g')\! \left( \frac{\kappa\nu l +g - g'}{\kappa (f+g)}
    \right)^{\!m}\!.
    \label{mean nth}
\end{equation}
Since the probability density $u(g)$ is normalized on the interval $[-g_{0},
g_{0}]$, from Eq.~(\ref{mean nth}) it follows that $\overline {\tau^{m}} \leq
(\nu l + 2g_{0}/\kappa )^{m}/ (f-g_{0})^{m}$. Thus, if $f>g_{0}$ then all these
moments are finite, and so in this case the classical central limit theorem for
sums of a random number of random variables \cite{GK} is applied to $Y_{t}$.
This implies that $\sigma^{2}(t) \propto t$ as $t \to \infty$, i.e., at
$f>g_{0}$ the biased diffusion of particles is normal, and the rescaled
probability density $\mathcal {P}(\psi, t) = \sigma(t) P(\langle Y_{t} \rangle
+ \sigma(t) \psi, t)$ in the long-time limit tends to the probability density
$\mathcal {P}(\psi, \infty) = (2\pi)^{-1/2}e^{-\psi^{2}/2}$ of the standard
normal distribution.

It is clear from the above that the anomalous long-time behavior of the
variance $\sigma^{2}(t)$ is expected at $\overline{\tau^{2}} = \infty$ when the
mentioned central limit theorem becomes inapplicable. The condition
$\overline{\tau^{2}} = \infty$ implies $f=g_{0}$ that, in accordance with
(\ref{tau}), yields $\tau_{\rm{min}} = \nu l/2g_{0}$ and $\tau_{\rm {max}} =
\infty$. Since the divergence of $\overline{\tau^{2}}$ occurs when $p(\tau)$ at
$\tau \to \infty$ tends to zero slowly enough, next we assume that
\begin{equation}
    p(\tau) \sim \frac{a}{\tau^{1+\alpha}}
    \label{asymp}
\end{equation}
$(\tau \to \infty)$ with $a>0$ and $\alpha \in (0,2]$. Thus, the biased
diffusion in a randomly layered medium is expected to be anomalous if both
conditions, $f=g_{0}$ and $\alpha \in (0,2]$, hold. The former guarantees that
the waiting time $\tau$ can be arbitrarily large ($\tau_{\rm{max}}= \infty$),
and so it is a necessary condition for anomalous diffusion. We note also that
in this case one may expect that the rescaled probability density $\mathcal
{P}(\psi, t)$ at $t\to \infty$ approaches the stable probability density, as
the generalized central limit theorem \cite{Feller} suggests.

The coefficient of proportionality $a$ and the exponent $\alpha$ are in general
not independent and can be found from the asymptotic behavior of the
probability density $u(g)$ in the vicinity of the point $g=-g_{0}$. Indeed,
using the condition $\int_{-g_{0}}^{g_{0}} dg\,gu(g) =0$, which is a
consequence of the symmetry property of $u(g)$, from Eq.~(\ref{p2}) at
$f=g_{0}$ we obtain
\begin{equation}
    p(\tau) \sim \frac{\nu l}{\tau^{2}} \left[ 1 - \frac{g_{0}}
    {\kappa \nu l}\! \left( 1 - \tau \frac{\partial}{\partial \tau}
    \right) \right] u\!\left(\frac{\nu l}{\tau} -g_{0}\right)
    \label{p3}
\end{equation}
($\tau \to \infty, \kappa \to \infty$). Then, assuming that
\begin{equation}
    u(g) \sim b \! \left( 1 + \frac{g}{g_{0}} \right)^{\! \beta -1}\!,
    \label{u as}
\end{equation}
where $g\to -g_{0}$, $b>0$ and $\beta>0$, the asymptotic formula (\ref{p3})
takes the form
\begin{equation}
    p(\tau) \sim bg_{0} \!\left( \frac{\nu l}{g_{0}} \right)^{\!\beta}
    \! \left( 1 - \frac{\beta g_{0}}{\kappa \nu l} \right)
    \! \frac{1}{\tau^{1+\beta}}.
    \label{p4}
\end{equation}
Finally, comparing (\ref{p4}) with (\ref{asymp}), we find $\alpha = \beta$ and
\begin{equation}
    a = a_{\rm{od}} \! \left( 1 - \frac{\alpha g_{0}}
    {\kappa \nu l} \right)\!,
    \label{a0}
\end{equation}
where $a_{\rm{od}} = bg_{0}(\nu l/g_{0})^{\alpha}$ is the parameter $a$ in the
overdamped limit. According to these results, the particle mass decreases the
parameter $a$ in comparison with the overdamped case but does not change the
exponent $\alpha$.

An example of $u(g)$ having the asymptotic behavior (\ref{u as}) is the
probability density
\begin{equation}
    u(g) = \frac{\Gamma(\beta + 1/2)}{g_{0}\sqrt{\pi}\,
    \Gamma(\beta)} \!\left( 1- \frac{g^{2}}{g_{0}^{2}}
    \right)^{\! \beta-1}\!,
    \label{u(g)}
\end{equation}
where $\Gamma(x) = \int_{0}^{\infty} dy\, y^{x-1} e^{-y}$ is the gamma
function, which corresponds to the symmetric beta distribution. According to
Eq.~(\ref{u(g)}), this distribution is unimodal with the maximum at $g=0$ if
$\beta>1$, bimodal with infinite maxima at $g= \pm g_{0}$ if $\beta<1$, and
uniform if $\beta=1$. Determining the parameter $b$ directly from the density
function (\ref{u(g)}), for this example we obtain
\begin{equation}
    a_{\rm{od}} = \frac{\Gamma(\alpha + 1/2)}{2\sqrt{\pi}\,
    \Gamma(\alpha)} \left( \frac{2\nu l}{g_{0}} \right)^{\!
    \alpha}.
    \label{a}
\end{equation}

\section{LAWS OF ANOMALOUS DIFFUSION}
\label{sec:Laws}

\subsection{Long-time behavior of the inverse Laplace transform}
\label{sec:Long}

From a formal point of view, the moments $\langle Y_{t} \rangle$ and $\langle
Y^{2}_{t} \rangle$ completely determine the variance $\sigma^{2}(t)$. But the
calculation of the inverse Laplace transforms in Eqs.~(\ref{1st}) and
(\ref{2nd}) is a difficult technical problem because of the contour integration
in the complex plane $s$. Fortunately, in the long-time limit this problem can
be avoided. Such a possibility provides the celebrated Tauberian theorem for
the Laplace transform \cite{Feller} which is widely used in the theory of CTRW
and its applications. According to this theorem, if $h(t)$ is ultimately
monotone and
\begin{equation}
    h_{s} = \mathcal{L}\{ h(t) \} \sim L\!\left( \frac{1}{s} \right)\!
    \frac{1}{s^{\rho}}
    \label{hs1}
\end{equation}
($0<\rho<\infty$) as $s \to 0$ then
\begin{equation}
    h(t) = \mathcal{L}^{-1}\{ h_{s} \} \sim \frac{1}{\Gamma(\rho)}\,
    L (t)\,t^{\rho-1}
    \label{ht1}
\end{equation}
as $t \to \infty$, where $L(t)$ is a slowly varying function at infinity. The
term `slowly varying' means that $L(\lambda t) \sim L(t)$, i.e., $\lim_{t\to
\infty} L(\lambda t) /L(t) =1$, for all $\lambda>0$. We note that, in contrast
to Eqs.~(\ref{1st}) and (\ref{2nd}), the parameter $s$ in (\ref{hs1}) is
assumed to be a positive real number.

The Tauberian theorem in the above form permits to find only the leading terms
of the asymptotic expansion of the moments $\langle Y_{t} \rangle$ and $\langle
Y^{2}_{t} \rangle$ as $t\to \infty$. If $\lim_{t\to \infty} \langle Y_{t}
\rangle^{2}/ \langle Y^{2}_{t} \rangle \neq 1$ then, according to the
definition (\ref{def var}), these terms completely determine also the leading
term of the long-time expansion of $\sigma^{2} (t)$. However, if $\lim_{t\to
\infty} \langle Y_{t} \rangle^{2}/ \langle Y^{2}_{t} \rangle = 1$ then for
finding the leading term of $\sigma^{2}(t)$ at least the first two terms of
each of the asymptotic expansions of $\langle Y_{t} \rangle$ and $\langle
Y^{2}_{t} \rangle$ should be evaluated. Remarkably, these terms can also be
determined from the Tauberian theorem if the Laplace transforms $\langle Y_{t}
\rangle_ {s}$ and $\langle Y^{2}_{t} \rangle_{s}$ at $s\to 0$ have the
asymptotic form
\begin{equation}
    h_{s} -  \frac{q}{s^{\eta}} \sim L\!\left( \frac{1}{s} \right)\!
    \frac{1}{s^{\rho}}
    \label{hs2}
\end{equation}
with $\eta>\rho$. In this case, replacing $h_{s}$ by $h_{s} - q/s^{\eta}$ and
using the exact result $\mathcal {L}^{-1} \{ 1/s^{\eta} \}=t^{\eta-1}/ \Gamma
(\eta)$ \cite{Erd}, from (\ref{ht1}) we obtain
\begin{equation}
    h(t) - \frac{q}{\Gamma(\eta)}t^{\eta-1} \sim \frac{1}{\Gamma(\rho)}\,
    L (t)\,t^{\rho-1}.
    \label{ht2}
\end{equation}

It should be noted that $qt^{\eta-1}/\Gamma(\eta)$ and $L (t)t^{\rho-1}
/\Gamma(\rho)$ actually represent the first two terms of the long-time
expansion of $h(t)$ only if $\eta>\rho$. In the opposite case, when $L(1/s)/
s^{\rho}$ is the leading term of the asymptotic expansion of $h_{s}$, this may
not be true. The reason is that in this case the second term of the asymptotic
expansion of $\mathcal {L}^{-1} \{ L(1/s)/s^{\rho} \}$ at $t\to \infty$ may not
be negligible in comparison with $qt^{\eta-1}/\Gamma(\eta)$. To illustrate this
fact, let us assume that $L(1/s) = k\ln(1/s)$ ($k$ is a scale factor) and $\rho
=2$, and consider the Laplace transform $h_{s} = q/s^{\eta} + (k/s^{2})\ln
(1/s)$. Since $\mathcal {L}^{-1} \{ (1/s^{2})\ln (1/s) \} = t(\ln t + \gamma
-1)$, where $\gamma = 0.5772$ is the Euler constant \cite{Erd}, for the inverse
Laplace transform of $h_{s}$ we obtain an exact result $h(t) = qt^{\eta-1}/
\Gamma(\eta) + kt(\ln t + \gamma -1)$. If $\eta>2$ and $t\to \infty$ then,
keeping in $h(t)$ the two leading terms, in accordance with (\ref{ht2}) we find
$h(t) - qt^{\eta-1}/ \Gamma(\eta) \sim kt \ln t$. However, if $\eta \leq 2$
then the second term of the asymptotic expansion of $\mathcal {L}^{-1} \{
L(1/s)/s^{\rho} \}$, $k(\gamma - 1)t$, is not negligible compared to $qt^{\eta-
1}/ \Gamma(\eta)$ and, as a consequence, the asymptotic formula (\ref{ht2})
does not hold. In this case only the leading term of $h(t)$, $kt\ln t$, is
determined from the Tauberian theorem. Thus, while at $\eta>\rho$ the first two
terms of the asymptotic expansion of $h(t)$ can be determined from the modified
Tauberian theorem, Eqs.~(\ref{hs2}) and (\ref{ht2}), to solve this problem in
the opposite case it is necessary to go beyond the Tauberian theorem.

Next, we use the Tauberian theorem, Eqs.~(\ref{hs1}) and (\ref{ht1}), and its
modified version, Eqs.~(\ref{hs2}) and (\ref{ht2}), to find the long-time
behavior of the first two moments, $\langle Y_{t} \rangle$ and $\langle
Y^{2}_{t} \rangle$, and the variance $\sigma^{2}(t)$ in the case of anomalous
diffusion, i.e., when the conditions $f=g_{0}$ and $\alpha \in (0,2]$ hold
simultaneously. Since the asymptotic solution of the CTRW is different for
different intervals of $\alpha$ \cite{Shles}, we consider the cases with
$\alpha \in (0,1)$, $\alpha \in (1,2)$, $\alpha =1$, and $\alpha =2$
separately.

\subsection{\mbox{\boldmath $\alpha\in (0,1)$}}

In this case it is convenient to represent the Laplace transform of the waiting
time probability density $p({\tau})$ in the form
\begin{equation}
    p_{s} = 1- \int_{\tau_{\rm{min}}}^{\infty} d\tau\,
    (1-e^{-s\tau})p(\tau),
    \label{def1}
\end{equation}
which follows from the definition $p_{s}= \int_{\tau_{\rm{min}}} ^{\infty}
d\tau\, e^{-s\tau}p(\tau)$ and the normalization condition $\int_{\tau_{
\rm{min}}} ^{\infty} d\tau\, p(\tau)=1$. Introducing the new variable of
integration $x=s\tau$ and using the asymptotic formula (\ref{asymp}), we obtain
\begin{equation}
    1-p_{s} \sim s^{\alpha} a \int_{0}^{\infty}dx\, \frac{1-e^{-x}}
    {x^{1+\alpha}}
    \label{as def1}
\end{equation}
as $s \to 0$. An integration by parts together with the integral representation
of the gamma function \cite{AS}, $\Gamma(x) = \int_{0}^{\infty}dy\, e^{-y} y^{x
- 1}$, reduces (\ref{as def1}) to the form
\begin{equation}
    1-p_{s} \sim \frac{a\Gamma(1-\alpha)}{\alpha}\,s^{\alpha}.
    \label{ps1}
\end{equation}

Now, using this result and the Laplace transforms
\begin{equation}
    \langle Y_{t} \rangle_{s} = l\, \frac{p_{s}}{s(1-p_{s})}, \quad\;
    \langle Y_{t}^{2} \rangle_{s} = l^{2} \frac{p_{s}^{2} + p_{s}}
    {s (1-p_{s})^{2}}
    \label{Y1,2s}
\end{equation}
of the first two moments of $Y_{t}$, we find in the limit $s\to 0$:
\begin{equation}
    \langle Y_{t} \rangle_{s} \sim \frac{l\alpha}{a\Gamma(1-\alpha)}\,
    \frac{1}{s^{1+\alpha}}
    \label{<X>s1}
\end{equation}
and
\begin{equation}
    \langle Y_{t}^{2} \rangle_{s} \sim \frac{2l^{2}\alpha^{2}}
    {a^{2}\Gamma^{2}(1-\alpha)}\,\frac{1}{s^{1+2\alpha}}.
    \label{<X2>s1}
\end{equation}
Since these asymptotic formulas are particular cases of the asymptotic formula
(\ref{hs1}) in which the slowly varying function $L(1/s)$ is a constant, from
(\ref{ht1}) we obtain in the long-time limit
\begin{equation}
    \langle Y_{t} \rangle \sim \frac{l\alpha}{a\Gamma(1-\alpha)
    \Gamma(1+\alpha)}\,t^{\alpha}
    \label{<X>1}
\end{equation}
and
\begin{equation}
    \langle Y_{t}^{2} \rangle \sim \frac{2l^{2}\alpha^{2}}
    {a^{2}\Gamma^{2}(1-\alpha)\Gamma(1+2\alpha)}\,t^{2\alpha}.
    \label{<X2>1}
\end{equation}
Thus, in this case $\lim_{t\to \infty} \langle Y_{t} \rangle^{2}/ \langle
Y^{2}_{t} \rangle \neq 1$ and the above asymptotic expressions yield
\begin{equation}
    \sigma^{2}(t) \sim \frac{l^{2}\alpha^{2}}{a^{2}\Gamma^{2}
    (1-\alpha)} \left(\! \frac{2} {\Gamma(1+2\alpha)} - \frac{1}
    {\Gamma^{2}(1+\alpha)}\right)\! t^{2\alpha}.
    \label{var1}
\end{equation}

According to this result, which agrees with that obtained in the context of the
asymptotic solution of the CTRW \cite{Shles}, subdiffusion occurs if $\alpha\in
(0,1/2)$ and superdiffusion if $\alpha\in (1/2,1)$. If $\alpha=1/2$ then
$\sigma^{2}(t) \propto t$ and, in accordance with the commonly used
terminology, the biased diffusion is normal. However, for normal diffusion
processes both the mean and variance are proportional to time. Therefore, since
$\langle Y_{t} \rangle \propto t^{1/2}$ at $\alpha = 1/2$, this type of
diffusion should be more appropriately termed as quasi-normal. It is also
worthy to note that, according to Eqs.~(\ref{a0}) and (\ref{var1}), the larger
is the particle mass, the stronger is diffusion.

\subsection{\mbox{\boldmath $\alpha\in (1,2)$}}

Since in this case $\lim_{t\to \infty} \langle Y_{t} \rangle^{2}/ \langle
Y^{2}_{t} \rangle = 1$ (see below), for finding the long-time behavior of
$\sigma^{2}(t)$ we should determine the first two terms of the asymptotic
expansion of $\langle Y_{t} \rangle$ and $\langle Y_{t}^{2} \rangle$ as $t\to
\infty$. To this end, taking into account that at $\alpha\in (1,2)$ the mean
waiting time $\overline{\tau} = \int_{\tau_{\rm{min}}}^ {\infty} d\tau\, \tau
p(\tau)$ exists, we use the following formula:
\begin{equation}
    p_{s} = 1- \overline{\tau}s - \int_{\tau_{\rm{min}}}^{\infty}
    d\tau (1-s\tau-e^{-s\tau}) p(\tau).
    \label{def2}
\end{equation}
Proceeding in the same way as before, we obtain
\begin{equation}
    p_{s} - 1 + \overline{\tau}s \sim \frac{a\Gamma(2-\alpha)}
    {\alpha(\alpha - 1)}\,s^{\alpha}
    \label{ps2}
\end{equation}
as $s\to 0$, and the straightforward calculation of the Lap\-lace transforms
(\ref{Y1,2s}) yields
\begin{equation}
    \langle Y_{t} \rangle_{s} - \frac{l}{\overline{\tau}}\,
    \frac{1}{s^{2}} \sim \frac{la\Gamma(2-\alpha)}
    {\overline{\tau}^{2}\alpha(\alpha -1)}\,
    \frac{1}{s^{3-\alpha}}
    \label{<X>s2}
\end{equation}
and
\begin{equation}
    \langle Y_{t}^{2} \rangle_{s} - \frac{2l^{2}}{\overline{\tau}^{2}}
    \,\frac{1}{s^{3}} \sim \frac{4l^{2}a\Gamma(2-\alpha)}
    {\overline{\tau}^{3}\alpha(\alpha -1)}\,\frac{1}{s^{4-\alpha}}.
    \label{<X2>s2}
\end{equation}

These asymptotic formulas are particular cases of the asymptotic formula
(\ref{hs2}) with $L(1/s)=\rm{const}$. Therefore, using the well-known property
of the gamma function, $\Gamma(1+x) = x\Gamma(x)$, from (\ref{ht2}) we get
\begin{equation}
    \langle Y_{t} \rangle - \frac{l}{\overline{\tau}}\,t \sim
    \frac{la}{\overline{\tau}^{2}\alpha(\alpha -1)(2-\alpha)}\,
    t^{2-\alpha}
    \label{<X>2}
\end{equation}
and
\begin{equation}
    \langle Y_{t}^{2} \rangle - \frac{l^{2}}{\overline{\tau}^{2}}
    \,t^{2} \sim \frac{4l^{2}a}{\overline{\tau}^{3}\alpha(\alpha -1)
    (2-\alpha)(3-\alpha)}\, t^{3-\alpha}
    \label{<X2>2}
\end{equation}
as $t\to \infty$. Accordingly, the long-time behavior of the variance
$\sigma^{2}(t)$ is described by the power law
\begin{equation}
    \sigma^{2}(t) \sim \frac{2l^{2}a}{\overline{\tau}^{3}\alpha
    (2-\alpha)(3-\alpha)}\, t^{3-\alpha}.
    \label{var2}
\end{equation}

Thus, since $\alpha \in (1,2)$, the transport of particles is superdiffusive.
Interestingly, depending on the exponent $\alpha$, the increase of the particle
mass $\mu$ can either enhance or suppress the biased diffusion. In order to
show this, we first use Eq.~(\ref{mean nth}) with $m=1$ to represent the mean
waiting time $\overline{\tau}$ in the form
\begin{equation}
    \overline{\tau} = \overline{\tau}_{\rm{od}}\! \left( 1 +
    \frac{1}{\kappa \overline{\tau}_{\rm{od}}} - \frac{g_{0}}
    {\kappa \nu l}\right)\!,
    \label{tau1}
\end{equation}
where $\overline{\tau}_{\rm{od}} = \nu l\int_{-g_{0}}^{g_{0}} dg\, u(g)/( g_{0}
+g)$ is the mean waiting time in the overdamped limit. Then, using this formula
for $\overline{\tau}$, the expression (\ref{a0}) for the parameter $a$ and the
condition $\kappa \tau_{\rm{min}} \gg 1$, we obtain
\begin{equation}
    \frac{a}{\overline{\tau}^{3}} = \frac{a_{\rm{od}}}
    {\overline{\tau}_{\rm{od}}^{3}} \! \left( 1 - \frac{3}
    {\kappa \overline{\tau}_{\rm{od}}} + \frac{(3-\alpha)g_{0}}
    {\kappa \nu l}\right)\!.
    \label{a/tau}
\end{equation}
According to this result, the biased diffusion is enhanced by the particle mass
at $(3-\alpha) \overline{\tau}_{ \rm{od}}>3\nu l/ g_{0}$ and is suppressed at
$(3-\alpha) \overline{\tau}_{ \rm{od}}<3\nu l/ g_{0}$. In particular, if the
probability density $u(g)$ is given by Eq.~(\ref{u(g)}) then
\begin{equation}
    \overline{\tau}_{\rm{od}} = \frac{\nu l}{g_{0}}\,
    \frac{\alpha -1/2}{\alpha -1},
    \label{mean tau}
\end{equation}
and so the former case occurs at $\alpha \in (1,3/2)$ and the latter at $\alpha
\in (3/2,2)$.

\subsection{\mbox{\boldmath $\alpha=1$}}

Here our starting point is the Laplace transform of the waiting time
probability density $p(\tau)$ represented as
\begin{equation}
    p_{s} = 1- aq_{s} - \int_{\tau_{\rm{min}}}^{\infty} d\tau
    (1-e^{-s\tau}) \left( p(\tau) - \frac{a}{\tau^{2}} \right)\!,
    \label{def3}
\end{equation}
where $q_{s} = \int_{\tau_{\rm{min}}}^{\infty} d\tau\, (1-e^{-s\tau})/
\tau^{2}$. The advantage of this representation is that the term $aq_{s}$
accounts for the asymptotic behavior of $p(\tau)$, $p(\tau) \sim a/\tau^{2}$ as
$\tau \to \infty$, in an explicit form. With the definition of the exponential
integral \cite{AS}, $E_{1}(x) =\int_{x}^{\infty} dy\,e^{-y}/y$, $q_{s}$ can be
written as
\begin{equation}
    q_{s} = s \frac{1-e^{-\xi}}{\xi} + sE_{1}(\xi)
    \label{qs1}
\end{equation}
($\xi=s\tau_{\rm{min}}$). Since $p(\tau) - a/\tau^{2} = o(1/\tau^{2})$ ($\tau
\to \infty$), the integral term in Eq.~(\ref{def3}) at $s\to 0$ can be
neglected compared to $aq_{s}$. Therefore, taking into account the asymptotic
formula $E_{1}(\xi) \sim \ln (1/s)$ \cite{AS}, we obtain
\begin{equation}
    1-p_{s} \sim as\ln \frac{1}{s}.
    \label{ps3}
\end{equation}
Using this result and Eq.~(\ref{Y1,2s}) for calculating the leading terms of
the Laplace transforms at $s\to 0$,
\begin{equation}
    \langle Y_{t} \rangle_{s} \sim  \frac{l}{a}\,\frac{1}{s^{2}
    \ln(1/s)}, \quad \;
    \langle Y_{t}^{2} \rangle_{s} \sim \frac{l^2}{a^2}\,
    \frac{2}{s^{3}\ln^{2}(1/s)},
    \label{X(1,2)s3}
\end{equation}
from the Tauberian theorem, Eqs.~(\ref{hs1}) and (\ref{ht1}), we find the
long-time behavior of the first two moments
\begin{equation}
    \langle Y_{t} \rangle \sim  \frac{l}{a}\,\frac{t}{\ln t},
    \quad \;
    \langle Y_{t}^{2} \rangle \sim \frac{l^2}{a^2}\,
    \frac{t^2}{\ln^{2} t}.
    \label{X(1,2)3}
\end{equation}

It should be noted that similar asymptotic formulas for $\langle Y_{t} \rangle$
and $\langle Y_{t}^{2} \rangle$ were obtained in Ref.~\cite{Shles}. But because
of the use of the waiting time probability density $p(\tau)$ of a particular
form, the asymptotic formulas derived in that paper do not depend on the
parameter $a$. At the same time, as it was shown above, the parameter $a$
contains an important information about the role of quenched disorder and
particle mass. In particular, Eq.~(\ref{a0}) shows that the moments
(\ref{X(1,2)3}) increase with the particle mass.

Since $\lim_{t\to \infty} \langle Y_{t} \rangle^{2}/ \langle Y_{t}^{2} \rangle
=1$, for finding $\sigma^{2}(t)$ as $t\to \infty$ we need to know at least the
two leading terms of the long-time expansion of $\langle Y_{t} \rangle$ and
$\langle Y_{t}^{2} \rangle$. In principle, using Eq.~(\ref{qs1}) and the
integral term in Eq.~(\ref{def3}), we could easily find the asymptotic behavior
of $1-p_{s} - as \ln(1/s)$ as $s\to 0$ and, in this way, obtain the second
terms of the asymptotic expansion of $\langle Y_{t} \rangle_{s}$ and $\langle
Y_{t}^{2} \rangle_{s}$. However, in contrast to the previous case, a
straightforward application of the Tauberian theorem to this case does not
provide a precise determination of the second terms of the asymptotic expansion
of $\langle Y_{t} \rangle$ and $\langle Y_{t}^{2} \rangle$. As it was argued in
Sec.~\ref{sec:Long}, in order to find these terms it is necessary to go beyond
the Tauberian theorem.

\subsection{\mbox{\boldmath $\alpha=2$}}

In this case we use the following representation for the Laplace transform of
$p(\tau)$:
\begin{equation}
    p_{s} = 1- \overline{\tau}s - ar_{s} - \int_{\tau_{\rm{min}}}^
    {\infty} d\tau (1-s\tau - e^{-s\tau}) \left( p(\tau) -
    \frac{a}{\tau^{3}} \right)\!,
    \label{def4}
\end{equation}
where
\begin{eqnarray}
    r_{s} &=& \int_{\tau_{\rm{min}}}^{\infty} d\tau\,
    \frac{1}{\tau^{3}}(1-s\tau - e^{-s\tau})
    \nonumber\\[6pt]
    &=& s^{2}\frac{1-2\xi - (1-\xi)\,
    e^{-\xi}}{2\xi^{2}} -\frac{s^{2}}{2}E_{1}(\xi).
    \label{rs}
\end{eqnarray}
This form of $p_{s}$ explicitly accounts for both the finiteness of $\overline
{\tau}$ and the asymptotic behavior of $p(\tau)$. Since $p(\tau) - a/\tau^{3} =
o(1/\tau^{3})$ ($\tau \to \infty$), at $s\to 0$ we can neglect the integral
term in Eq.~(\ref{def4}) in comparison with $ar_{s}$. This, together with the
asymptotic formula $r_{s} \sim -(s^{2}/2) \ln(1/s)$ ($s\to 0$), yields
\begin{equation}
    p_{s} - 1 + \overline{\tau}s \sim \frac{a}{2}s^{2}\ln \frac{1}{s}.
    \label{ps4}
\end{equation}

Using this result, the Laplace transforms of the first two moments of $Y_{t}$
at $s\to 0$ can be written as
\begin{equation}
    \langle Y_{t} \rangle_{s} - \frac{l}{\overline{\tau}}\,
    \frac{1}{s^{2}} \sim \frac{la} {2\overline{\tau}^{2}}\,
    \frac{1}{s}\ln \frac{1}{s}
    \label{X(1)s4}
\end{equation}
and
\begin{equation}
    \langle Y_{t}^{2} \rangle_{s} - \frac{2l^{2}}
    {\overline{\tau}^{2}} \,\frac{1}{s^{3}} \sim \frac{2l^{2}a}
    {\overline{\tau}^{3}}\,\frac{1}{s^{2}} \ln \frac{1}{s}.
    \label{X(2)s4}
\end{equation}
Therefore, in accordance with the modified Tauberian theorem, Eqs.~(\ref{hs2})
and (\ref{ht2}), we obtain
\begin{equation}
    \langle Y_{t} \rangle - \frac{l}{\overline
    {\tau}}\, t \sim \frac{la}{2\overline{\tau}^{2}}\,\ln t
    \label{X(1)4}
\end{equation}
and
\begin{equation}
    \langle Y_{t}^{2} \rangle - \frac{l^{2}}{\overline{\tau}^{2}}
    \,t^{2} \sim \frac{2l^{2}a} {\overline{\tau}^{3}}\, t\ln t.
    \label{X(2)4}
\end{equation}
As a consequence, the long-time behavior of the variance is described by the
asymptotic formula
\begin{equation}
    \sigma^{2}(t) \sim \frac{l^{2}a} {\overline{\tau}^{3}}\, t\ln t.
    \label{var4}
\end{equation}
The fact that $\sigma^{2}(t)$ increases faster than $t$ is in accordance with
the asymptotic formula (\ref{var2}). Indeed, while $t^{3-\alpha}$ approaches
$t$, the coefficient of proportionality between $\sigma^{2}(t)$ and $t$ tends
to infinity as $\alpha \to 2-0$. We note also that the ratio $a/\overline
{\tau}^{3}$ at $\alpha = 2$ is determined by the same Eq.~(\ref{a/tau}).
Therefore, if $u(g)$ is given by Eq.~(\ref{u(g)}) then the larger is the
particle mass, the weaker is diffusion.

In conclusion of this subsection we would like to draw attention to the
differences between our model and one-dimensional iterated maps which generate
trajectories according to the rule $x_{n+1} = x_{n} + F(x_{n})$. It is usually
assumed \cite{DetMap} that $F(x)$ is an antisymmetric, $F(-x)=-F(x)$, and
periodic, $F(x+N)=F(x)$ ($N$ is an integer), function. Due to these conditions,
there is no drift, i.e., the quantity $\langle x \rangle \equiv \langle x_{n+t}
- x_{n} \rangle $ can be taken to be zero, where the angular brackets denote an
average over a properly chosen set of initial conditions of $x_{n}$ and $t$
plays the role of the number of iterations. The same property, $\langle x
\rangle =0$, holds also for maps perturbed by time dependent noise \cite{BMWG}
and quenched disorder \cite{Rad} with zero means. Thus, these maps are unbiased
and so the variance of $x_{n+t} - x_{n}$, i.e., $\langle x^{2} \rangle -
\langle x \rangle^{2}$, reduces to $\langle x^{2} \rangle$. Using the CTRW
theory, in most cases it is possible to write this variance as the inverse
Laplace transform and use the ordinary Tauberian theorem for finding its
long-time behavior \cite{DetMap, BMWG}. In contrast, in our model $\langle
Y_{t} \rangle$ grows with time and although the moments $\langle Y_{t} \rangle$
and $\langle Y_{t}^{2} \rangle$ can also be represented as the inverse Laplace
transform, the variance $\sigma^{2}(t) = \langle Y_{t}^{2} \rangle - \langle
Y_{t} \rangle^{2}$ cannot. Therefore, if $\lim_{t\to \infty} \langle Y_{t}
\rangle^{2}/ \langle Y_{t}^{2} \rangle =1$, the leading term of the asymptotic
expansion of $\sigma^{2}(t)$ as $t \to \infty$ cannot be determined by applying
the ordinary Tauberian theorem. We have solved this problem (for $\alpha \neq
1$) by using the modified Tauberian theorem. It should also be noted that
adding to $F(x)$ a weak uniform bias breaks the symmetry of the system and, as
a consequence, leads to a time dependence of $\langle x \rangle$
\cite{BiasMap}. However, the biased maps considered in \cite{BiasMap} do not
exhibit anomalous diffusion at long times.

\subsection{Role of thermal fluctuations}

We complete our analysis with a qualitative discussion of the role of thermal
fluctuations. These fluctuations can be accounted for by adding the thermal
noise term to the right-hand side of Eq.~(\ref{eq motion}). In this case some
important conclusions can be drawn from the asymptotic behavior of the
correlator $K(x,y) = \langle [U(x) - U(y)]^{2} \rangle$ as $|x - y|\to \infty$,
where $U(x)$ is the random potential that corresponds to the random force $g(x)
= -dU(x)/dx$. Since under thermal fluctuations particles can move in both
directions, the random force and potential should be determined on the entire
$x$-axis. Using the properties of $g(x)$ and the continuity condition for
$U(x)$ with $U(0) = 0$, we obtain
\begin{equation}
    U(x) = -(x - nl)g^{(n)} + U(nl).
    \label{U(x)}
\end{equation}
Here, $x \in [nl, nl+l)$, $n=0,\pm 1,\ldots$, and $U(nl) = -l \sum_{s=0}^{n}
g^{(s)} +lg^{(n)}$ if $n\geq 0$ and $U(nl) = l \sum_{s=-1}^{n} g^{(s)}$ if $n
\leq -1$. Taking into account that the random forces $g^{(s)}$ with different
$s$ are statistically independent, the definition of the correlator $K(x,y)$
and Eq.~(\ref{U(x)}) lead to the asymptotic expression
\begin{equation}
    K(x,y) \sim l \sigma^{2}_{g} |x-y|
    \label{K(x,y)}
\end{equation}
($|x-y| \to \infty$), where $\sigma^{2}_{g}= \int_{-g_{0}}^{g_{0}}dg \,g^{2}
u(g)$.

The systems with $K(x,y) \propto |x-y|$ have long been a subject of extensive
study (see, e.g., Refs.~\cite{BG,Mon} and references therein). A remarkable
result obtained for these systems in the overdamped regime is that there always
exists a threshold value $f_{\rm{tr}}$ of the external force $f$ in which the
depinning transition occurs. This transition is characterized by vanishing the
average particle velocity in the pinning state, when $f\leq f_{\rm{tr}}$, while
in the depinning state, when $f>f_{\rm{tr}}$, particles move with a nonzero
average velocity which strongly depends on $f$. According to \cite{Den}, if
$g(x)$ is a bounded function then $f_{\rm{tr}}<g_{0}$ ($f_{\rm{tr}}\to g_{0}$
as the temperature approaches zero) and the mean first-passage time in the
depinning and pinning states is finite and infinite, respectively. Since the
moments of the first-passage time can be associated with the moments of the
waiting time, we may expect therefore that at nonzero temperatures and
$f<g_{0}$ the exponent $\alpha$ and so the character of anomalous diffusion
becomes depending on $f$ (the diffusion behavior at $f>g_{0}$ is expected to be
normal). Specifically, with decreasing of $f$ from $g_{0}$ to $f_{\rm{tr}}$ the
exponent $\alpha$ should also decrease from 2 to 1, and if $f<f_{\rm{tr}}$ then
$\alpha<1$. Of course, in order to find the dependence of $\alpha$ on $f$ and
the diffusion laws a quantitative consideration of the problem is needed. It is
especially important because the effects arising from the joint action of
quenched disorder and thermal fluctuations are often unexpected and even
counterintuitive.

\section{CONCLUSIONS}
\label{sec:Concl}

We have studied in the long-time limit the unidirectional transport of
particles which occurs under a constant force in a randomly layered medium. The
influence of the layers is modeled by a piecewise constant random force whose
values in different layers are assumed to be independent and identically
distributed with bounded support. We have reduced the problem of the
unidirectional transport, initially formulated in the framework of the motion
equation, to a continuous-time random walk on a semi-infinite chain. The main
statistical characteristic of this approach, the waiting time probability
density, is expressed through the probability density of the random force and
particle characteristics, including the particle mass. By analyzing the
dependence of the moments of the waiting time on the external force, we have
formulated the conditions under which the biased diffusion exhibits the
anomalous behavior. It has been shown that this behavior may occur only if the
external force is equal to the boundary value of the random force.

In order to find in the anomalous regime the long-time behavior of the first
and second moments of the particle position, we have used the Tauberian theorem
and its modified version allowing, in most cases, to determine the first two
terms of the asymptotic expansion of these moments. Within this approach, we
have found, with one exception, the explicit asymptotic formulas for the
variance of the particle position, i.e., the laws of diffusion. The time
dependence of the variance is completely controlled by the exponent describing
the asymptotic behavior of the probability density of the random force in the
vicinity of its minimum value. It has also been shown that, depending on the
value of this exponent, the particle mass in the weakly underdamped regime can
either enhance or suppress the anomalous diffusion without changing its time
dependence.

\section*{ACKNOWLEDGMENT}

We are grateful to an anonymous referee for constructive criticism and helpful
suggestions.

\end{document}